\documentclass[preprint]{aastex62}

\usepackage{ulem}
\usepackage{xcolor}
\usepackage{graphicx}

\received{September 19, 2019}
\revised{November 8, 2019}
\accepted{November 20, 2019}
\submitjournal{\apj}

\shorttitle{First PSP results from QTN}
\shortauthors{Moncuquet et al.}

\begin{document}

\title{First in-situ Measurements of Electron Density and Temperature\\ from Quasi-Thermal Noise Spectroscopy with Parker Solar Probe/FIELDS}

\correspondingauthor{Michel Moncuquet}
\email{michel.moncuquet@obspm.fr}

\author[0000-0002-0786-7307]{Michel Moncuquet}
\affil{LESIA, Observatoire de Paris, Universit\'e PSL, CNRS, Sorbonne
Universit\'e, Universit\'e de Paris \\
5 place Jules Janssen, 
92195 Meudon, France}

\author[0000-0001-6449-5274]{Nicole Meyer-Vernet}
\affil{LESIA, Observatoire de Paris, Universit\'e PSL, CNRS, Sorbonne
Universit\'e, Universit\'e de Paris \\
5 place Jules Janssen, 
92195 Meudon, France}

\author[0000-0002-2757-101X]{Karine Issautier}
\affil{LESIA, Observatoire de Paris, Universit\'e PSL, CNRS, Sorbonne
Universit\'e, Universit\'e de Paris \\
5 place Jules Janssen, 92195 Meudon, France}

\author[0000-0002-1573-7457]{Marc Pulupa}
\affiliation{Space Sciences Laboratory, University of California, Berkeley, CA 94720-7450, USA}

\author[0000-0002-0675-7907]{J. W. Bonnell}
\affil{Space Sciences Laboratory, University of California, Berkeley, CA 94720-7450, USA}


\author[0000-0002-1989-3596]{Stuart D. Bale}
\affil{Space Sciences Laboratory, University of California, Berkeley, CA 94720-7450, USA}
\affil{Physics Department, University of California, Berkeley, CA 94720-7300, USA}
\affil{The Blackett Laboratory, Imperial College London, London, SW7 2AZ, UK}
\affil{School of Physics and Astronomy, Queen Mary University of London, London E1 4NS, UK}

\author[0000-0002-4401-0943]{Thierry {Dudok de Wit}}
\affil{LPC2E, CNRS and University of Orl\'eans, Orl\'eans, France}

\author[0000-0003-0420-3633]{Keith Goetz}
\affiliation{School of Physics and Astronomy, University of Minnesota, Minneapolis, MN 55455, USA}

\author[0000-0001-8956-2824]{L\'ea Griton}
\affil{IRAP, Universit\'e Toulouse III - Paul Sabatier, CNRS, CNES, Toulouse, France}
\affil{LESIA, Observatoire de Paris, Universit\'e PSL, CNRS, Sorbonne Universit\'e, Universit\'e de Paris \\
5 place Jules Janssen, 92195 Meudon, France}

\author[0000-0002-6938-0166]{Peter R. Harvey}
\affil{Space Sciences Laboratory, University of California, Berkeley, CA 94720-7450, USA}

\author[0000-0003-3112-4201]{Robert J. MacDowall}
\affil{Solar System Exploration Division, NASA/Goddard Space Flight Center, Greenbelt, MD, 20771, USA}

\author[0000-0001-6172-5062]{Milan Maksimovic}
\affil{LESIA, Observatoire de Paris, Universit\'e PSL, CNRS, Sorbonne Universit\'e, Universit\'e de Paris \\
5 place Jules Janssen, 92195 Meudon, France}

\author[0000-0003-1191-1558]{David M. Malaspina}
\affil{Laboratory for Atmospheric and Space Physics, University of Colorado, Boulder, CO 80303, USA}

\begin{abstract}

Heat transport in the solar corona and wind is still a  major unsolved astrophysical problem. Because of the key role played by electrons, the electron density and temperature(s) are  important  prerequisites for understanding these plasmas. We present such in situ measurements along the two first solar encounters of Parker Solar Probe (PSP), between 0.5 and 0.17 AU  from the Sun, revealing different states of the emerging solar wind near solar activity minimum. These preliminary results are obtained from a simplified analysis of the plasma quasi-thermal noise (QTN) spectrum measured by the Radio Frequency Spectrometer (RFS/FIELDS).  The local electron density is deduced from the tracking of the plasma line, which enables accurate measurements, independent of calibrations and spacecraft perturbations, whereas the temperatures of the thermal and supra-thermal components of the velocity distribution, as well as the average kinetic temperature are deduced from the shape of the plasma line.  The temperature of the weakly collisional thermal population, similar for both encounters,  decreases with distance as $R^{-0.74}$, much slower than adiabatic. In contrast, the temperature of the  nearly collisionless suprathermal population exhibits a virtually  flat radial variation. The 7-second resolution of the density  measurements enables us to deduce  the  low-frequency spectrum of compressive fluctuations around perihelion, varying as $f^{-1.4}$. This is the first time that  QTN spectroscopy  is implemented with an electric antenna length not exceeding the plasma Debye length. As PSP will approach the Sun, the decrease in Debye length is expected to considerably improve the accuracy of the temperature measurements.

\end{abstract}


\section{Introduction} \label{sec:intro}

The Parker Solar Probe (PSP) spacecraft \citep{fox16}, launched on August 12, 2018, is orbiting the Sun   on highly elliptical trajectories  of perihelion gradually    decreasing from 35.7 solar radii ($R_\sun$)  to a closest approach of 9.86 $R_\sun$ from the center of the Sun, via Venus gravity assists.  The present paper deals with the two first encounters that took place in October-November 2018 (E01) and March-April 2019 (E02), close to the heliographic equator, with perihelions of 0.17 AU (35.7 $R_\sun$),  therefore largely extending inwards the Helios exploration. 
{ The trajectory  crossed several times the heliospheric current sheet, thus revealing different types of wind and dynamic structures \citep{bal19}.}
During the so-called encounter phases of the trajectory, inward of 0.25 AU (54 $R_\sun$), all instruments record data at a high rate (7 second cadence). In order to increase the radial extension of the results, we have also analyzed data farther away from the Sun,  up to about 100 $R_\sun$ (0.46 AU), despite the reduced rate (56 second cadence). 

Our results  are based on power spectra acquired by the low-frequency receiver (LFR) of the Radio Frequency Spectrometer (RFS), part of the FIELDS instrument suite on PSP \citep{bal16}. The RFS instrument \citep{pul17} is a two-channel receiver and spectrometer, at the terminals of four 2-m monopole electric antennas mounted near the front of the spacecraft close to the extremities of the  heat  shield diagonals, so that the two  corresponding linear dipoles are of 7-m tip-to-tip length, perpendicular to the axis of the spacecraft. The present preliminary results are based on data from only one dipole (V$_1$-V$_2$). The LFR   (10.5 kHz - 1.7 MHz) spectra are analyzed with the technique of quasi-thermal noise (QTN) spectroscopy, which yields the electron density and the  temperatures of the thermal (core) and suprathermal components of the velocity distribution, as well as the total kinetic temperature (e.g. \citet{mey17} and references therein).

The radial temperature profiles have never been measured in situ inward of 0.3 AU.  Helios I and II - from 0.3 to 1 AU, and  Ulysses - from about 1 to 4  AU,  found approximate power-law decreases with distance of indices typically between  -0.3 and -0.9 for the thermal core, with a flatter gradient for fast streams \citep{sit80,pil90,phi95a,iss98,iss99a,mak00}. The suprathermal part,  which includes a beaming component  aligned to the magnetic field (also known as Strahl), was previously found to represent 4-10 \% of the distribution \citep{mcc92,stv09} and to have a temperature  decreasing weakly with distance  \citep{phi95b,pie16}.

{ Despite decades of study, the turbulence in the solar wind is still not fully understood, especially the role of compressive fluctuations in the transport of energy in a weakly collisional plasma. A question of considerable importance is the relationship between the small-scale structure associated with density fluctuations and large-scale plasma properties \citep{lion16, ale13,lacombe14}.   The accurate electron density samples obtained from QTN on PSP at perihelion enable us to study these compressive fluctuations much closer to the Sun than previously \citep{celni83,mar90,saf15}.}

{ The paper is organized as follow. Section 2 recalls the main properties of QTN under PSP conditions and gives complete radio spectrograms for both solar encounters. Section 3 presents the methods to deduce the electron density and temperatures with some analytical approximations. Section 4 shows the radial variations of the electron thermal and  suprathermal  temperatures  during the two first extended encounters, discusses the results and produces statistics.  Section 5 uses the electron density data sets around 35.7 $R_\sun$ to deduce the low-frequency spectrum of compressive fluctuations. Final remarks as conclusions are given in Section 6.}

\section{Quasi-Thermal Noise Spectroscopy \label{sec:qtn}}

Plasma particle properties in space are classically measured in situ by particle analyzers, pioneered in the solar wind by the  `solar plasma experiment' onboard Mariner 2 which provided,  more than half-a-century ago  \citep{neu62, neu97}  the ultimate proof that this supersonic wind was more than  a mere theoretician dream \citep{par58, par01}.

In contrast, the technique of QTN spectroscopy measures particles via electrostatic fields, exploiting the strong coupling between plasma particles  and fields { \citep{sit67}}. Introduced  onboard  ISEE 3 \citep{mey79}, { and pioneered to measure the cold \citep{mey86a} and hot \citep{mey86b} electrons in the tail of a comet,}  it uses the power spectrum of the voltage induced on an electric antenna by the particle quasi-thermal motions, measured by a radio receiver connected to an electric antenna. The signature of the electrons is a line at the electron plasma frequency $f_p$, which reveals the total electron density $n \propto f_p^2$, whereas the shape of the line reveals the electron kinetic temperature, as well as its  thermal (core) and suprathermal components  {  (\cite{mey17} and references therein).}

The $f_p$ plasma line is produced by Langmuir waves induced by the particle quasi-thermal motions. Since the Langmuir wavelength $\lambda_L$ exceeds the ambient Debye length $L_D$, the detection requires an electric antenna of length exceeding   $L_D$. { However, electrons interact with waves of phase speed equal to their proper speed, and since  $\lambda_L \rightarrow \infty$ as the frequency $f\rightarrow f_p$, so does the Langmuir wave phase speed. Hence suprathermal electrons can increase considerably  the spectral density  at $f_p$, producing a peak of amplitude characteristic of them \citep{mey89}. On the other hand, the electrons passing-by the antenna closer than $L_D$ induce transient voltages of duration $1/f_p$, which thus produce a flat spectrum for $f<f_p$, characteristic of the thermal core of the electron velocity distribution. Furthermore, the high-frequency spectrum is proportional to the total electron pressure.}

{ Because $\lambda_L \rightarrow \infty$ as $f \rightarrow  f_p$,} the technique is equivalent to a detector of large cross-section -  much larger than that of conventional space-borne detectors, and it is relatively immune to spacecraft perturbations, photoelectrons and charging effects \citep{mey98a}. For these reasons, QTN spectroscopy is complementary to particle analyzers, serving routinely to calibrate them  \citep [e.g.,][]{mak95, iss01,sal01}, and has been and will be implemented on a number of spacecraft in various environments (see e.g. \cite{mon09}).

\begin{figure}[ht!]
\vspace*{-30mm}
\plottwo{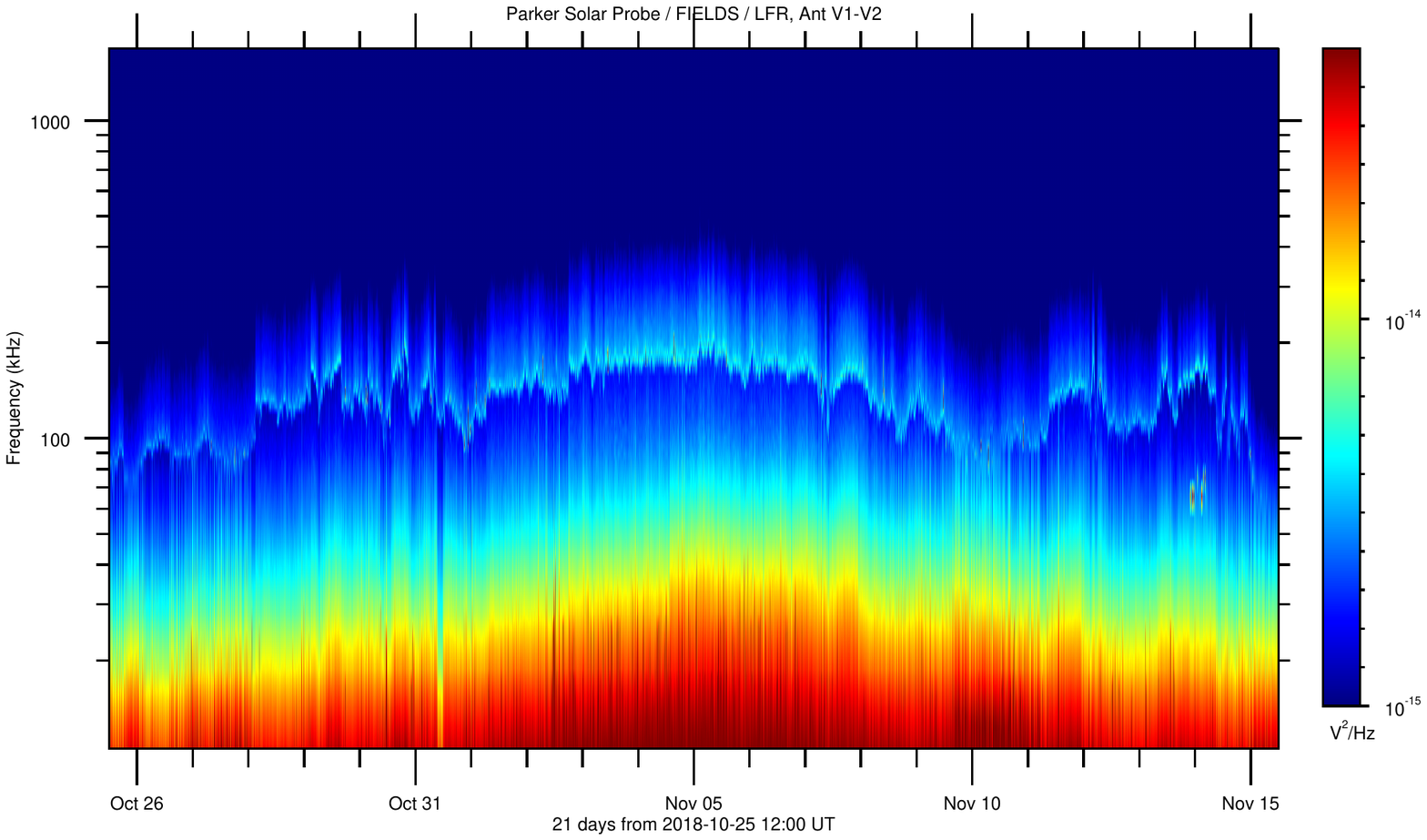}{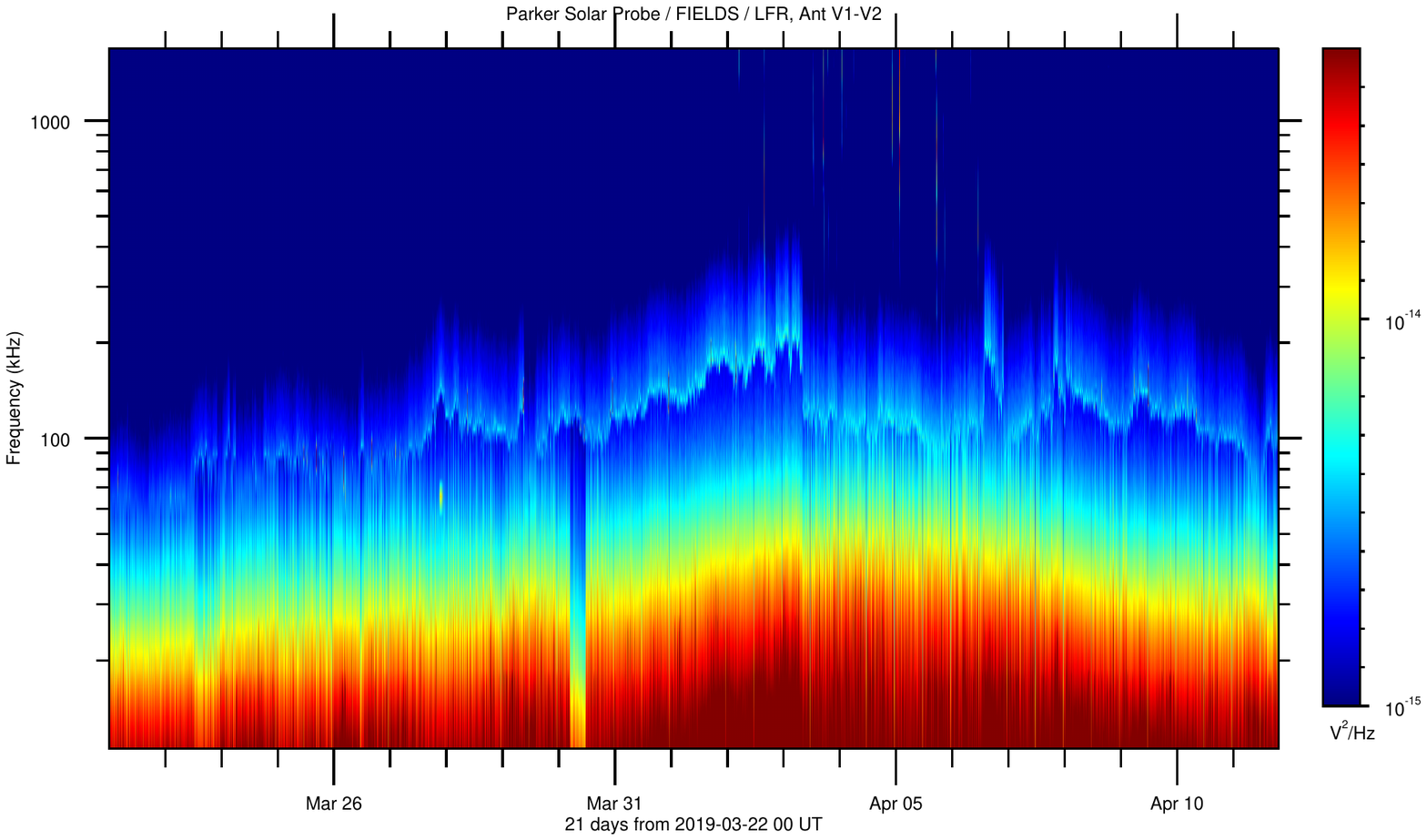}

\vspace*{-30mm}
\caption{\label{spectre} 21-days spectrograms acquired during the first (2018, October 25 - November 15, left panel) and second (2019, March 22 - April 11, right panel) PSP solar encounters, showing  the plasma quasi-thermal noise on which the plasma line at $f_p$  emerges clearly (cyan line varying between 80 and 200 kHz). }
\end{figure}

These properties are especially suitable on PSP because near the Sun, the 2-m electric antennas will be adequately longer than $L_D$, whereas the expected complex  environment of the spacecraft \citep{erg10} will require a technique immune to spacecraft perturbations in order to measure the genuine plasma particle properties. However, the perihelion of the first PSP  orbits, lying  outwards of $35 R_\sun$, is not close enough to the Sun for the Debye length to be smaller than the antenna  length, which was unfortunately restricted by spacecraft safety considerations. Nevertheless, the  supra-thermal electrons,  of speed close to the phase speed of Langmuir waves near $f_p$, enables the plasma line to emerge (Figure  \ref{spectre}).

\section{Deducing the electron density and temperatures \label{sec:param}} 

\begin{figure}[ht!]
\plotone{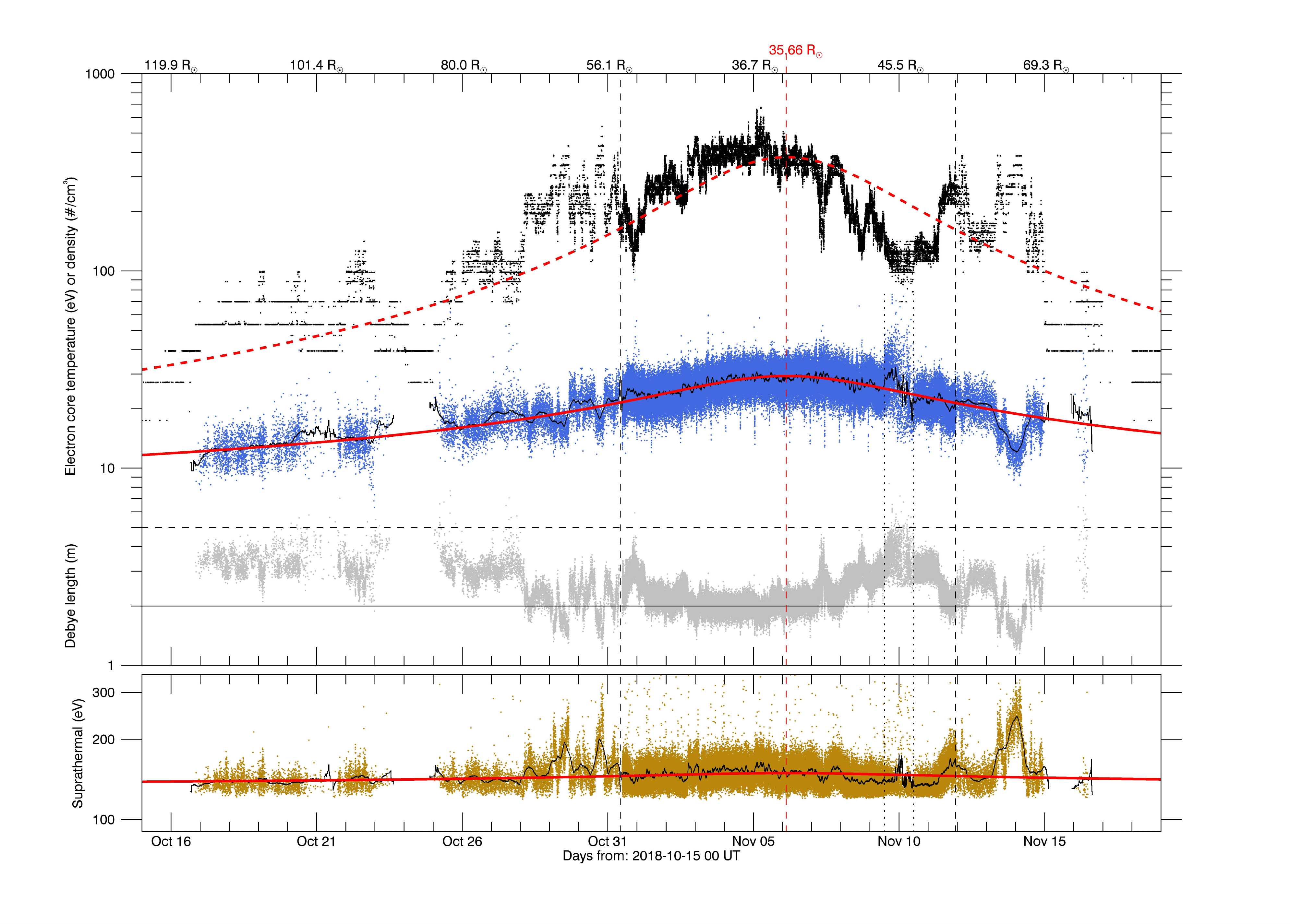}

 \vspace*{-7mm}

\plotone{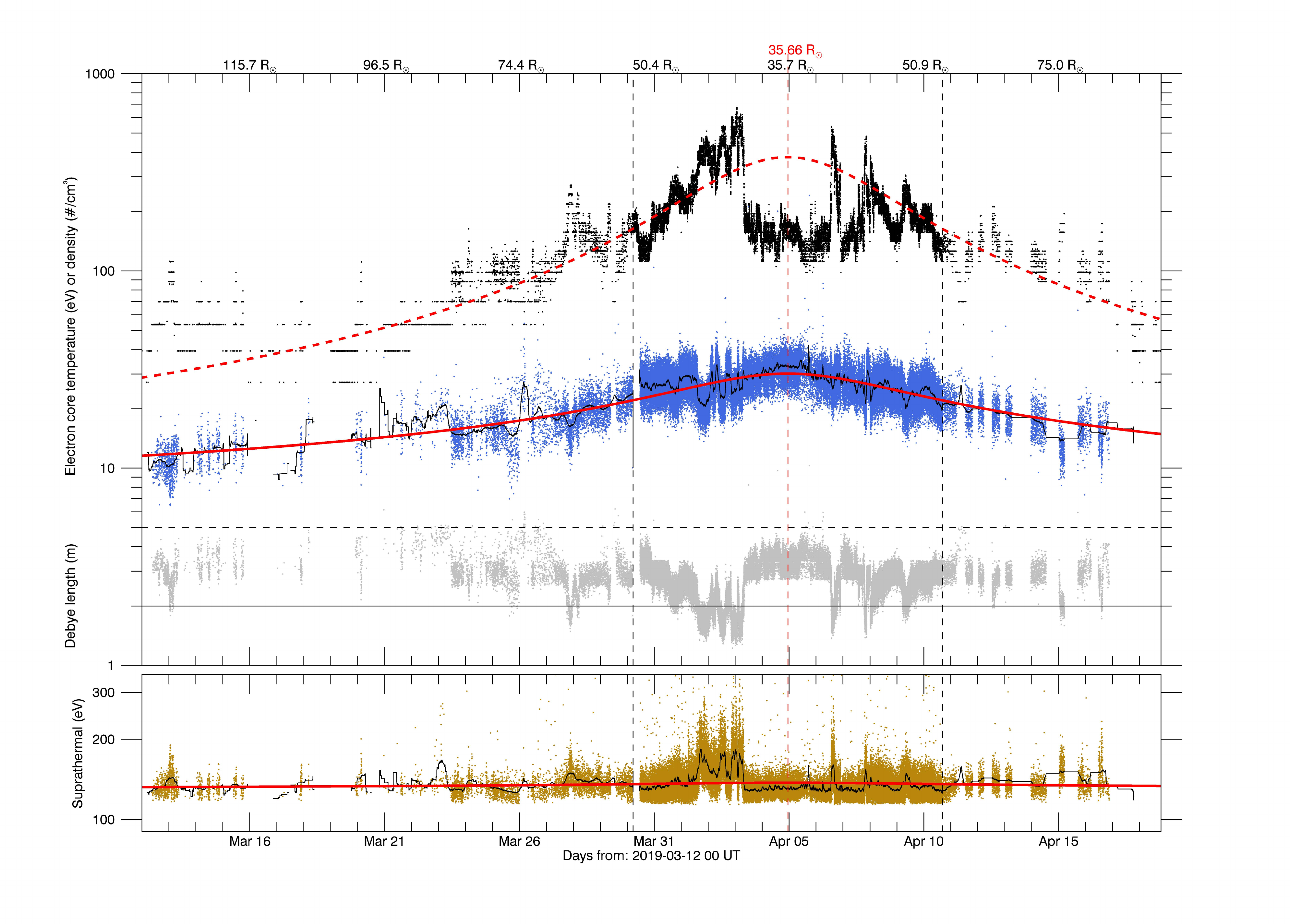}
       
       \vspace*{-10mm}
	\caption{\label{nTc} {\footnotesize Electron total density (black, in cm$^{-3}$, with an arbitrary $10\times R_{AU}^{-2}$ variation superimposed in red) and  temperatures (in eV, with the fitted radial variations shown in Figure \ref{profiles} superimposed in red) of the thermal  (core, in blue) and suprathermal (in gold) components of the electron velocity distribution for the first (2018, October 15 - November 18, top panel) and second  (2019, March 12 - April 18, bottom panel) extended encounters. The solid black line over the $T_c$ and $T_h$ values is a one-hour smoothing. The corresponding plasma Debye length (m) is plotted in grey ( compared to the 2-m antenna length in black). The heliocentric distance (in Sun radius) is indicated at the top of each panel (in red for each PSP perihelion) .}}
\end{figure}

The electron density and temperatures are usually obtained from the QTN technique by assuming a model for the velocity distribution, calculating the theoretical QTN power spectral density, and  deducing the parameters of the model by fitting the theory to the observations (e.g. \citet{iss98}). However for the  present preliminary results, we used a simpler  method, based on  simplified estimates of the relevant parts of the spectrum, similar to the method used on Cassini in Saturn's magnetosphere \citep{mon05, sch13}, albeit at Saturn, the  Debye length was short compared to the antenna length, whereas the opposite is true for these two first encounters (see Figure \ref{nTc}).  Instead of using the whole spectra  in the LFR-RFS frequency range  with model fitting, we determine the electron density from the frequency of the detected peak, and the temperatures of thermal and suprathermal components of the electron distribution from the power level reached at the peak and  the minimum level below the peak.

{
\subsection{Electron density}    }
First of all, the local electron density is deduced from the tracking of the plasma line at $f_p$ { (see an example in Figure \ref{gudule})}, with elimination of questionable measurements .  Miscellaneous algorithms may be used for tracking the plasma line from the raw power spectrum at the inputs of the receiver, without any calibration, and a software to do so is implemented in RFS/FIELDS \citep{pul17}. In the present study, we use instead the algorithm developed for the SORBET radio receiver  \citep{mon06, kasab19} on the BepiColombo (ESA-JAXA) mission, which is mainly based on detecting the steepest growth rate in each raw spectrum. This algorithm (somewhat improved for LFR-RSF) is efficient; for example, during the high-rate data period in the first encounter (E01), the peak is detected in 95\% of the available spectra ( $\sim$ 130000 spectra). The main improvement of the algorithm was to withdraw the  false positives (for example with type III bursts or strong interferences). We used the fact that a supposed high $f_p$ must provide a relative high peak level if it is {\it bona-fide} QTN  (since the peak  behaves roughly as $\sqrt f_p$, cf. Eq.(\ref{eqthk}) below); otherwise  the detection is withdrawn: this represents about 2\% of the initial selection (3\% for the second encounter, likely because of the numerous type III bursts and the dilute wind around perihelion).

The deduced electron density is  plotted (black dots) on Figure \ref{nTc}. Because they stem from a detection algorithm (instead of a model fitting)  these results  are stepped since  LFR-RFS uses only 64 pseudo-logarithmically spaced frequencies in the range 10 kHz -1.7 MHz. As a consequence, the lowest  densities  (say $<100 $cm$^{-3}$) are more stepped than the higher ones, with larger error bars. 

{
\subsection{Electron temperatures} }
\begin{enumerate}

\item 
The temperature $T_c$ of the thermal component of the  velocity distribution, assumed Maxwellian, is then deduced from the voltage spectral density just below $f_p$,  hereafter noted $V^2_{min}$ { (see Figure \ref{gudule})}. This quantity is essentially the plateau QTN spectrum, hereafter noted $V^2_{0}$,  produced by electrons passing around the antennas closer than $L_D$  on which the effect of the separation of the antenna arms is discussed in the Appendix A.  
To calculate $V^2_{0}$ at the receiver inputs, one must take into account the antenna gain calculated from the load/stray capacitance and the impedance of the antenna modified by the plasma. The QTN plateau is then given as a function of the core temperature $T_c$ and Debye length $L_D$ by :  (from \cite{mon05}) 
\begin{equation}
V^2_{0} \approx  \frac{ 8\sqrt{2 m_e k_B T_c}}{\pi^{3/2}\varepsilon_0 (1+C_B/C_A)^2}  \int_0^\infty \frac{F(kL) k L_D^2 }{[k^2L_D^2+1]^2}dk 
\label{v2min}
\end{equation}
in S.I. units,  $V^2_{0}$ being in ${\rm V^2/Hz}$. Here $F(kL)$ is the PSP wire antenna response { (detailed in appendix A)}, with $L$ the single wire length ($L \simeq 2 $m), $C_A={\pi \varepsilon_{0} L}/{\ln(L_{D}/a)}$ is an approximation of the dipole antenna capacitance at low frequencies, with $a$ the wire radius ($a \simeq $1.5 mm), and $C_B$ is the (dipole) stray capacitance ($\sim 18$ pF).
{ To this plateau QTN must be added two (generally minor) contributions : the shot noise $V^2_{shot}$,  produced by the currents flowing between the antennas and the plasma, possibly mitigated/enhanced by the antenna biasing, and the Doppler-shifted protons QTN $V^2_{p}$. Finally, we  use an iterative method to solve the implicit Eq.(\ref{v2min}) and deduce $T_c$ from the observable $V^2_{min}$ , taking into account at each step some approximations of  $V^2_{shot}$ and $V^2_{p}$ (see Appendix B).}

The deduced  core temperature $T_c$   during the first (top) and second (bottom) extended solar encounters is plotted in blue on Figure \ref{nTc}. The vertical dashed lines indicate the beginning and end of the high-rate data (7-sec). The vertical dotted lines indicate a period when the data are perturbed by biasing or change in modes; this is unfortunate because this day is one of the rare periods of high wind speed in E01, therefore preventing us to measure the thermal electron properties in high speed wind. 

\item
The temperature of the suprathermal component of the velocity distribution is obtained using the ratio ${V^2_{max}}/{V^2_{min}}$ between the peak level  and the level $V^2_{min}$ { (see Figure  \ref{gudule})}, and is therefore independent of calibration or biasing which affect the two levels in the same way. Statistically, this ratio is well centered with regular variance. Modelling the suprathermal electrons by a Kappa distribution (which includes the Maxwellian case $\kappa \rightarrow\infty$), the peak level may be estimated using Eqs.(59)-(60) by  \citet{mey17} (with range of validity therein, including the case $L_D>L$).   We deduce the following estimate for the  temperature $T_{h}$ of the suprathermal electrons:
\begin{equation}
\frac{V^2_{max}}{V^2_{min}} \approx \frac{\kappa-3/2}{\kappa} \frac{F(A \; L/L_D)}{F_0(L/L_D)} \sqrt{\frac{\pi}{12} \frac{f_p}{\Delta f} }  \frac{T_{h}}{\sqrt{T_c T}} 
\label{eqthk}
\end{equation}
where  $\Delta f$ is the frequency resolution of the instrument at the plasma frequency $f_p$, $T$ the electron kinetic temperature, $A \simeq \sqrt{T_c/T} \sqrt{2 \Delta f / 3f_p }$  and $F_0$ the value of the integral in Eq.(\ref{v2min}) (also detailed in Appendix A).

Given the uncertainties on the whole electron velocity distribution and the approximations made in our QTN modelling, which presently does not take into account the anisotropy of the Strahl, we will only exploit the centered estimator  ${V^2_{max}}/{V^2_{min}}$ obtained from the data, without trying to match a value of kappa.  In other words, we deduce a proxy $T_h$ of the temperature of the suprathermal component of the distribution assuming merely $T_h \propto {V^2_{max}}/{V^2_{min}}$, and using Eq.(\ref{eqthk}) to dimensionate, with $\kappa = 5$. These  temperatures are plotted in gold at the bottom of Figure \ref{nTc}  for the first (top) and second (bottom) extended solar encounters. Note that the results for different values of kappa can be deduced from (\ref{eqthk}), for example with Maxwellians $T_h$ would be 30\% smaller.

\item
The  electron kinetic temperature $T$ is deduced from the high-frequency part of the spectrum. For antennas long enough with respect to $L_D$, the high-frequency QTN varies as $f^{-3}$ and is directly proportional to the kinetic temperature and independent of the separation between the antenna arms \citep{mey89} - a result independent of the velocity distribution provided it is isotropic \citep{cha91}. However, for the first and second PSP orbits, the Debye length is not small enough, so that the kinetic temperature can be estimated from fitting the high-frequency part of the spectrum, after subtraction of the galactic noise and the receiver noise \citep{mak19}. Note that, contrary to the determination of $T_c$, such a simplified method cannot be implemented in the presence of electromagnetic emissions, which were frequent during encounter E02   \citep{pul19}; it is also  much less  accurate than if obtained by fitting the whole spectrum, since at high frequencies the QTN is of the same order of magnitude as the  receiver noise and the galactic noise.  
 
\end{enumerate}

\section{Large scale variations \label{sec:variations}} 

\subsection{Radial profiles of electron temperatures \label{sec:profiles}} 

\begin{figure}[ht!]
\plotone{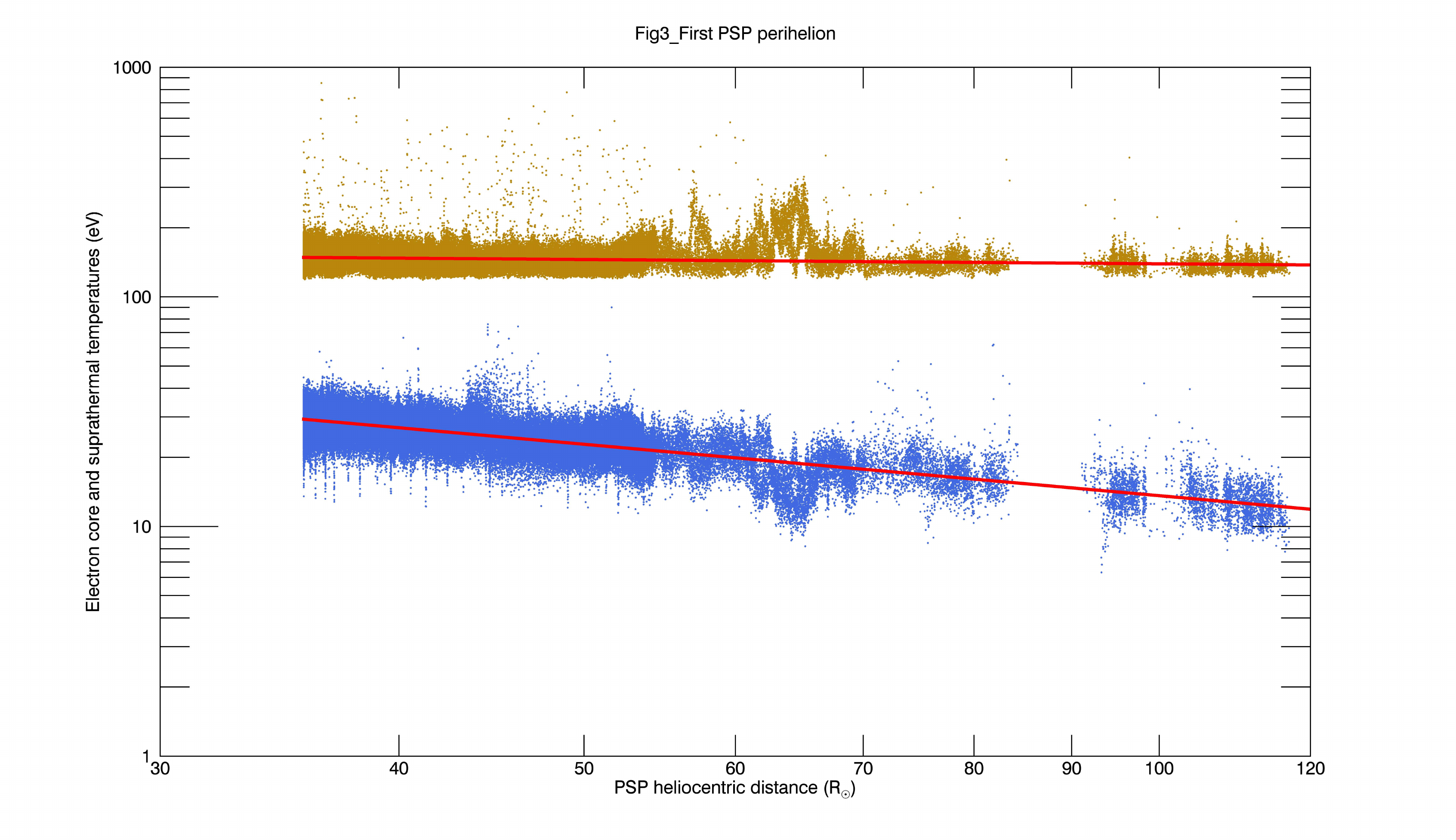}
\plotone{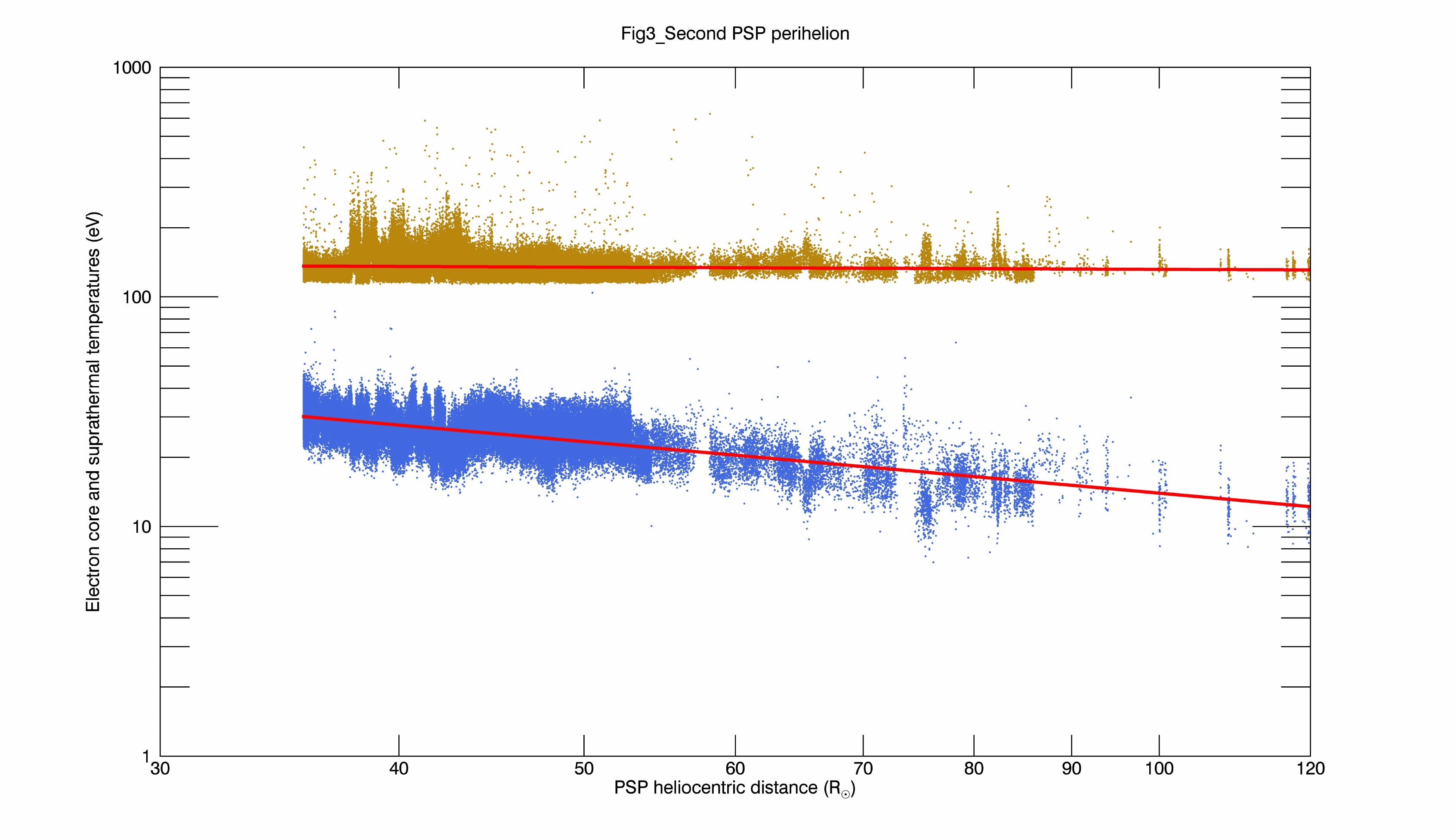}
\caption{\label{profiles} Variation with heliocentric distance of the temperatures of the thermal (core, in blue) and suprathermal (in gold) components of the electron velocity distribution for the extended encounters 1 (top) and 2 (bottom), with the fitted profiles superimposed  in red.}
\end{figure}

Figure~\ref{profiles} shows the thermal (core, in blue) and  suprathermal (in gold) electron temperatures as a function of heliocentric distance $R$ in units of solar radius $R_\sun$, with the fitted power laws (in red) given by
\begin{eqnarray}
T_{c\textrm{(eV)}}  & \simeq & 418 \times (R/{R_\sun})^{-0.74}  \label{Fit1} \;\;\;\;\; T_{h\textrm{(eV)}}   \simeq  185 \times (R/{R_\sun})^{-0.06}   \;\;\;\;\;                      \textrm{for E01} \label{Tc1} \\
T_{c\textrm{(eV)}}  & \simeq & 432 \times (R/{R_\sun})^{-0.74}  \label{Fit2} \;\;\;\;\; T_{h\textrm{(eV)}}   \simeq  153 \times (R/{R_\sun})^{-0.03}   \;\;\;\;\;  \textrm{for E02} \label{Tc2}
\end{eqnarray}

for respectively the first and second extended encounters. Note that these results are obtained using robust straight-line fits , i.e. minimizing the mean absolute deviation:  from its value, we  estimate an error on the logarithmic slopes of the profiles of about $\pm 0.03$.

Both extended encounters have similar temperature profiles despite the differences in densities. The thermal temperature is about $3.5 \times 10^5 $ K at 36 $R_\sun$ (0.17 AU), and $2.3 \times 10^5$ K at 64  $R_\sun$ (0.3 AU). The latter value and the radial  profile are similar to those found by Helios \citep{pil90,mak05}, despite the general decrease in solar activity till the Helios epoch, and the expected overall decrease in coronal temperature \citep{sch14}. In contrast, the radial profile of the  suprathermal electron temperature is nearly flat. The flatness of $T_h$ agrees with the  Helios results inward of 1 AU \citep{mak05,pie16}, but the absolute values are somewhat higher   (1.6  million degrees at 64  $R_\sun$ (0.3 AU)). This might be due to the crudeness of our $T_h$ measurements (see Section \ref{sec:param}), for which the logarithmic slope of the  $T_h$ profile is more accurate than the absolute values. However, extrapolating to 1.5 $R_\sun$ yields a temperature of $1.7-2.1 \times 10^6$ K, which is in the range of electron temperatures measured in the quiet corona \citep{dav98} and in streamers at solar activity minimum \citep{koh97,gib99}.

{
\subsection{Discussion}  }
With the density and core temperature measured at 0.17 AU, the mean free path of thermal electrons is of the same order of magnitude of the pressure scale height, i.e. their Knudsen number   is of order of magnitude unity. Under such weakly collisional conditions, the heat flux is not given by the Spitzer-H\"{a}rm value, whose validity requires  a much smaller  Knudsen number \citep{scu79,sho83}, but collisions are not negligible, requiring numerical kinetic simulations. Such simulations, taking into account Coulomb collisions and spherical expansion  with a radial magnetic field, the electrostatic field produced by the electron-proton mass difference being computed self-consistently,  yield a thermal electron temperature decreasing with distance with a logarithmic slope in the range  $0.6 - 0.9$ and an antisunward drifting suprathermal component, tending to have an isothermal profile  \citep{lan12}.

This nearly isothermal behavior of suprathermal electrons, with a  nearly constant relative density of a few times $(m_e/m_p)^{1/2}$ ($m_e$ and $m_p$ being the electron and proton masses) were  predicted  by \citet{mey98b} using an analytical approximation of an exospheric solar wind  model \citep{lem71} with  a radial magnetic field. The constant  kinetic temperature of suprathermal electrons is a simple consequence of conservation of energy and magnetic momentum for electrons escaping from the solar electrostatic potential, and is consistent with an adiabatic anisotropic fluid behavior, whereas the value of the relative density comes from the equality of electron and proton densities and escaping fluxes. The suitability of exospheric models for  suprathermal electrons is due to  the increase in Coulomb  free path  as the square of the energy, making them collisionless. This assumes that if  instabilities \citep{mar06} and scattering by turbulent fluctuations at electronic scales change the velocity directions, they  change negligibly { the kinetic energy per particle} of this component of the distribution. However, as expected, this agreement does not hold for the collisional thermal electrons, for which exospheric theory predicts an adiabatic radial variation with logarithmic slope $-4/3$, as a consequence of their trapping by the solar electrostatic potential  \citep{mey03}. The weaker observed slope is presumably due to the heat flux carried by the suprathermal electrons, which  decreases radially faster than $R^{-2}$ \citep{sci01}, implying heat deposition in the plasma. 
A more detailed comparison should take into account that our measurements of $T_c$ and $T_h$ concern the temperatures of the two untruncated components of the velocity distribution, whereas the exospheric distinction between populations involves truncations in velocity space.

Finally, from the relation $T=T_c+(n_h/n) \times T_h$, where $T$ is the   electron kinetic temperature, $T_c$ and $T_h$ those of respectively the thermal (core) and suprathermal  component of the distribution, and  $n_h/n$ the relative density of the suprathermals, assumed to be about 10\% from Helios measurements  \citep{stv09}, we obtain $T \simeq 5 \times 10^5$ K at 0.17 AU,  in agreement with the high frequency QTN \citep{mak19}.

\subsection{\label{sec:histograms} Statistics of the electron density and core temperature during E01 and E02}

{ Most of the large-scale differences in density between E01 and E02 apparent on Figure \ref{nTc} do not seem to have counterparts in the electron thermal temperatures. To examine this point, we use below our large data set to study the different regimes encountered.}

\begin{figure}[ht!]
    \plottwo{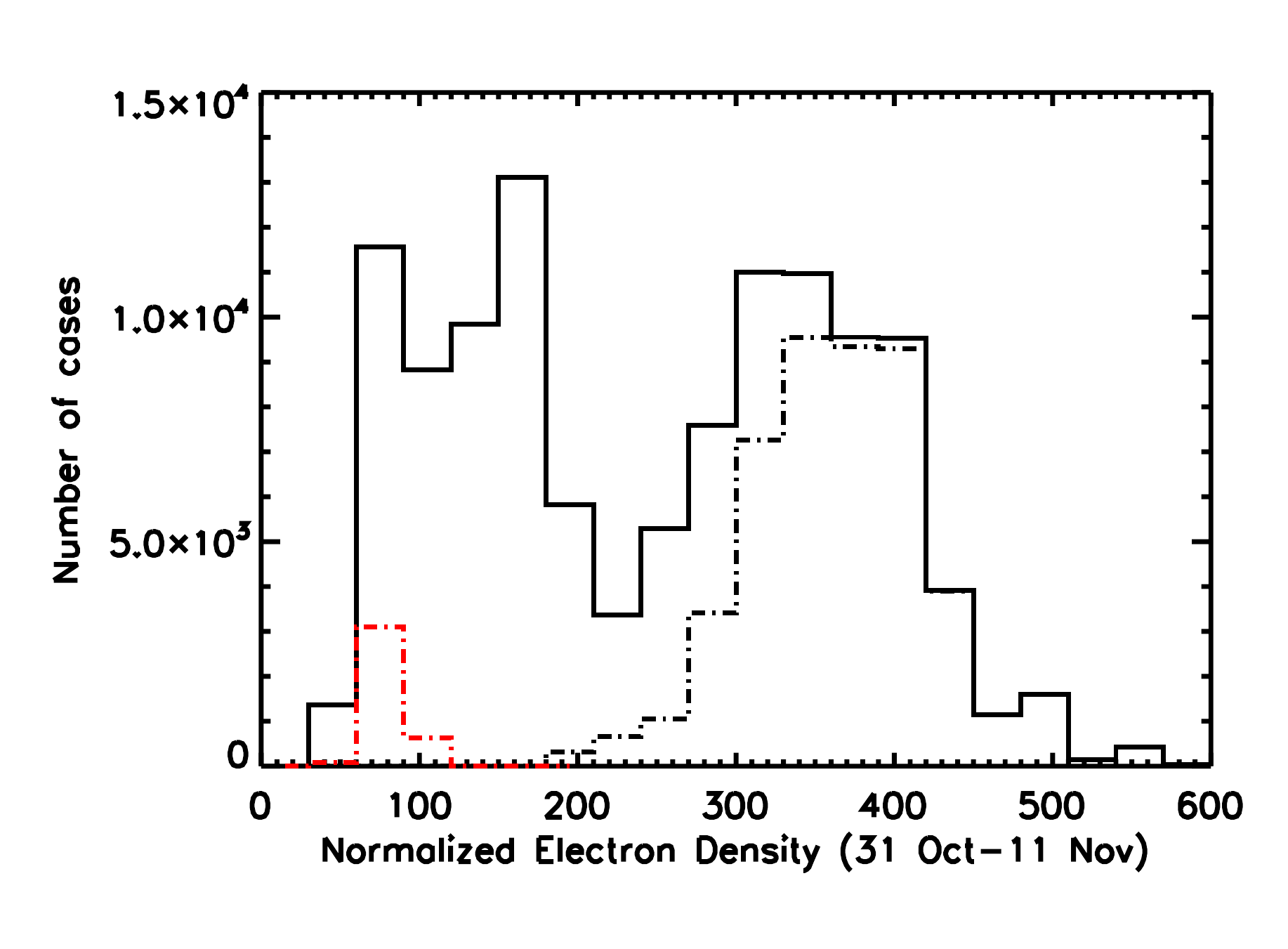}{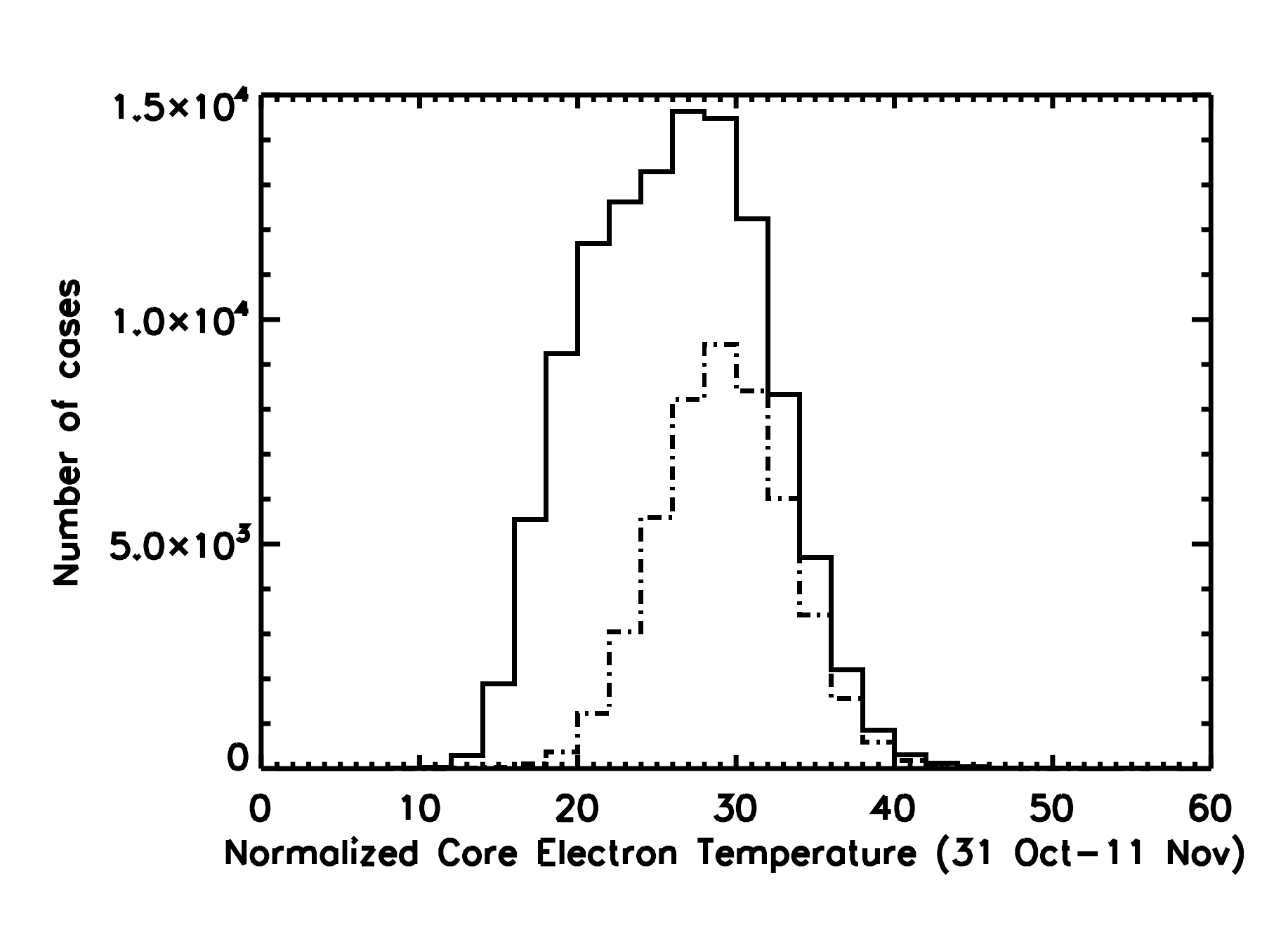}
    \plottwo{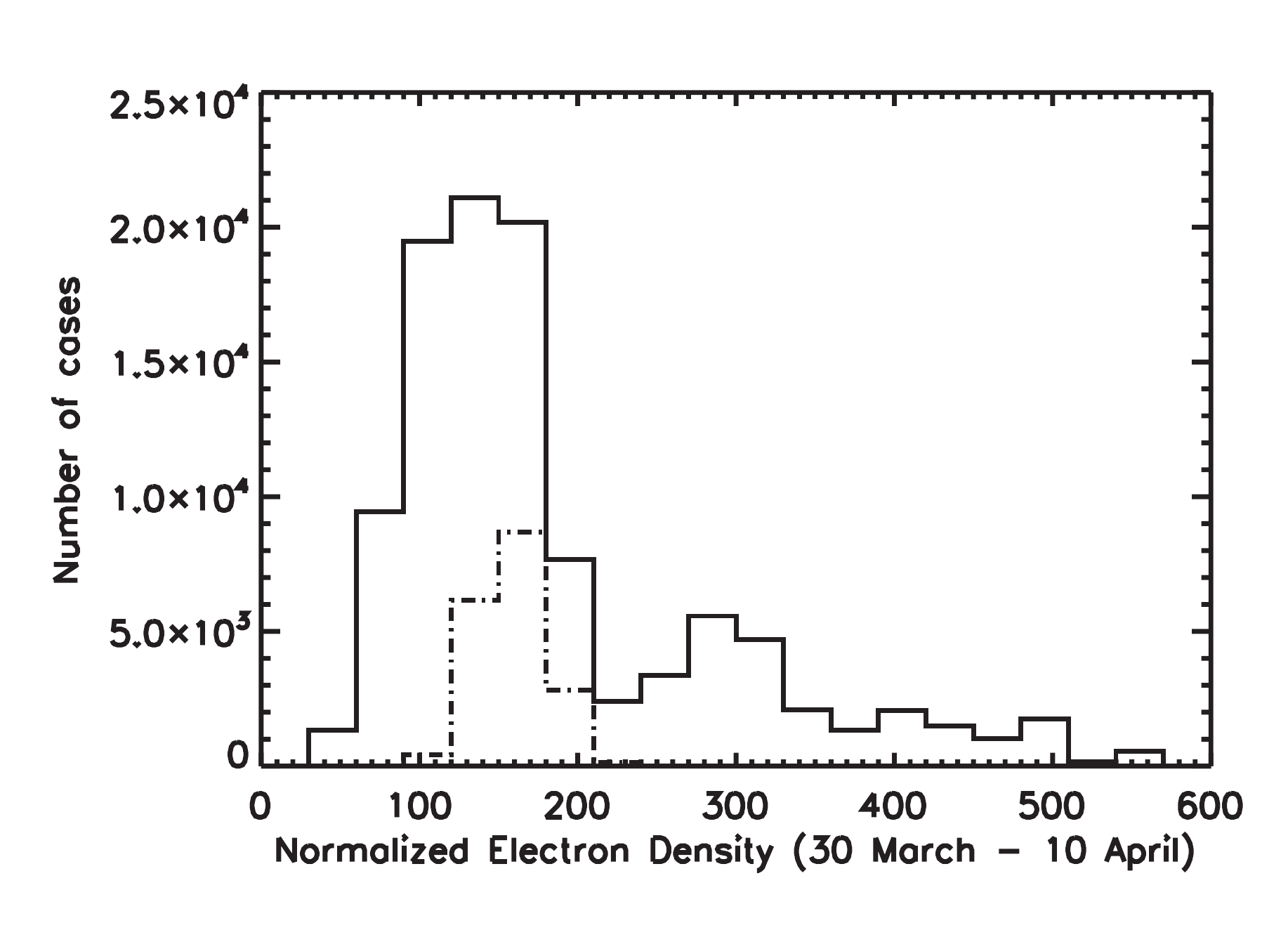}{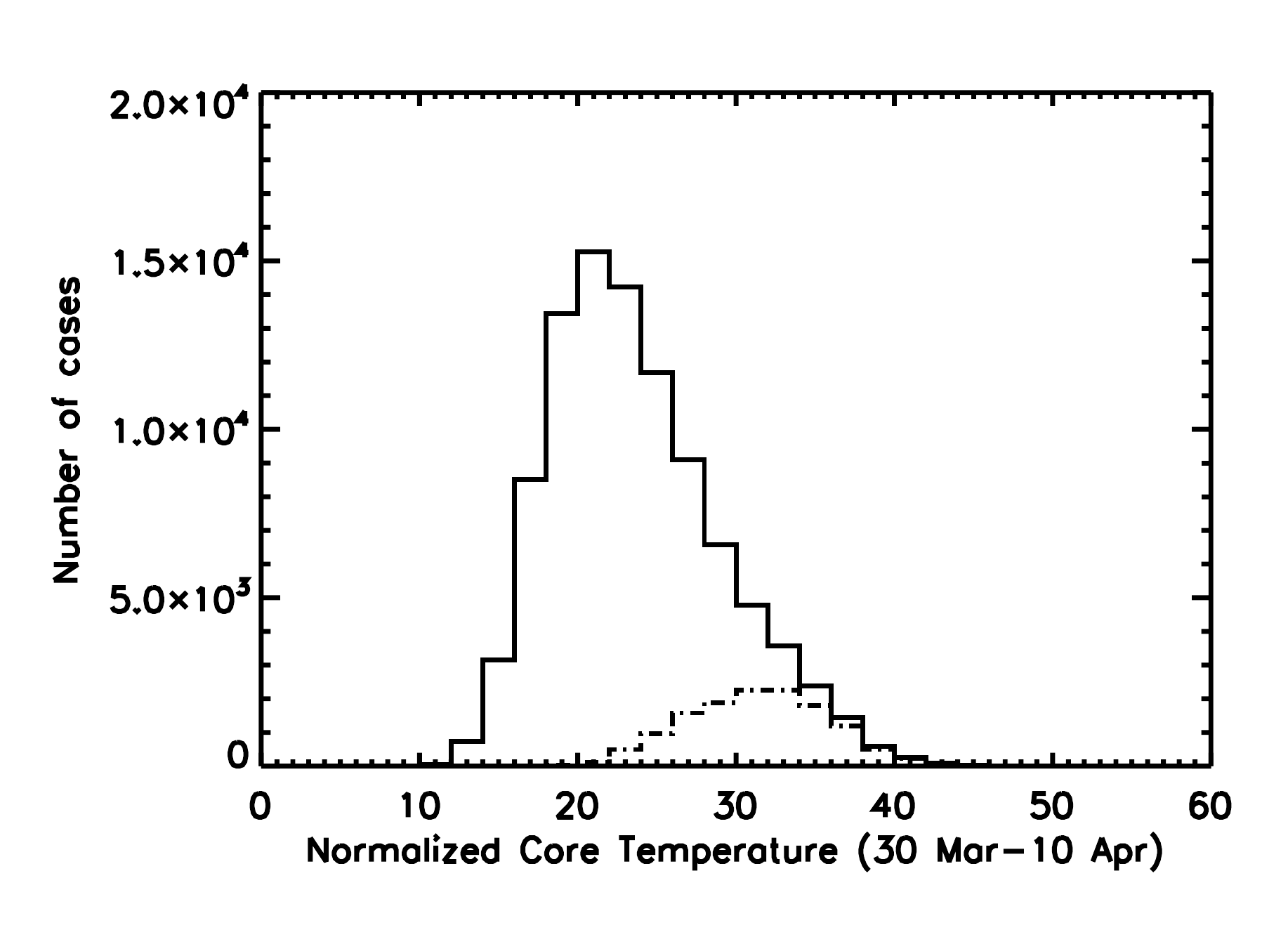}
    \caption{\label{histogramsN} Histograms of the electron density (left) and the core (thermal) electron temperature (right) for both encounters. For E01, we superimpose in the top panels a dilute wind period (9-10 nov) as a red dotted-dash line and a period of dense wind (grey distribution) around the first perihelion (3-6 nov). For E02, on bottom panels, we over plot as a dotted-dash grey line the histogram of density and temperature measured around the second perihelion  (3-6 April 2019).}
\end{figure}

The top panels of Figure \ref{histogramsN} represent the histograms of the electron density ($\sim$ 115,000 data points) and core electron temperature ($\sim$ 112,500 data points) during the first encounter E01, i.e., 31 October to 11 November 2018, with a 7-sec resolution (solid line histograms). We normalize each histogram to 0.17 AU, using an $R^{-2}$ density dependence and the $R^{-0.74}$ power-law determined in section \ref{sec:variations} for $T_c$. 
During the approach to E01, PSP showed a complex solar wind structure with different regimes, i.e., slow and intermediate wind from streamers, flow interactions, in addition to sporadic faster flows from small equatorial coronal holes, which are typical of equatorial regions \citep{phi95c,iss98, neug2001, iss05}.
For E01, the electron density histogram exhibits two main kinds of populations: a dilute electron population (with a mean value $\simeq$ 130 cm$^{-3}$ at 35.7 $R_\sun$) and a dense electron population (with a mean value  $\simeq$ 330 cm$^{-3}$). We superimpose a grey dotted-dash histogram obtained during E01, from 3 to 6 November 2018: we do see that this period mainly contains a denser wind population with a mean value of $\sim$370 cm$^{-3}$. The red dotted-dash histogram is obtained on midday 9 to 11 November 2018, where a faster solar wind is measured \citep{kasp19}. We can thus attribute the most dilute electron distribution to the fastest wind, in agreement with the well-known anticorrelation between the solar wind density and speed \citep{neug2001}. 
In contrast, the shape of the core electron temperature histogram shows mainly one single population for E01. It has a mean value of 26 eV whereas the grey dotted-dash core temperature has 
a mean around 29 eV. Contrary to the electron density, the core temperature is more difficult to associate to specific structures. In particular, many interplanetary events from denser density populations could be correlated with the same kind of temperature distributions \citep{sal03}.

We have  also plotted the distributions of the scaled electron density ($\sim$ 120,000 data points) and core temperature ($\sim$ 105,000 data points) at 0.17 AU for E02, on the bottom panels of Figure \ref{histogramsN}. We normalize each histogram to 0.17 AU as we did for E01. The density populations are likely to be associated with 3 major classes of solar wind:  the quiet undisturbed wind with lower mean values of density $\simeq$ 120 cm$^{-3}$, which dominates for E02; the denser heliospheric plasma sheet with intermediate values of density ($\simeq$ 290 cm$^{-3}$); the over dense disturbed wind with interplanetary shocks, density compressions regions, etc., with higher mean values of density ($\simeq$ 400 cm$^{-3}$). Note that these mean values correspond to those obtained at 1 AU, shifted to 35.7 $R_\sun$. The grey dotted-dash histogram is obtained from 3 to 6 April 2019, where the density is 50 \% smaller than the one obtained during E01, i.e., $\simeq$ 160 cm$^{-3}$. The corresponding core temperature has  a mean value around 31.5 eV. We can conclude that for this period the dilute density is correlated to the highest core electron temperature. Future works will need a detailed study on the large-scale structure of the wind, using in particular the wind speed data. 
\\

\section{Low-frequency compressive turbulence in the pristine solar wind \label{sec:turbulence}}

\begin{figure}[ht!]
  \centering
    \includegraphics[scale=0.60]{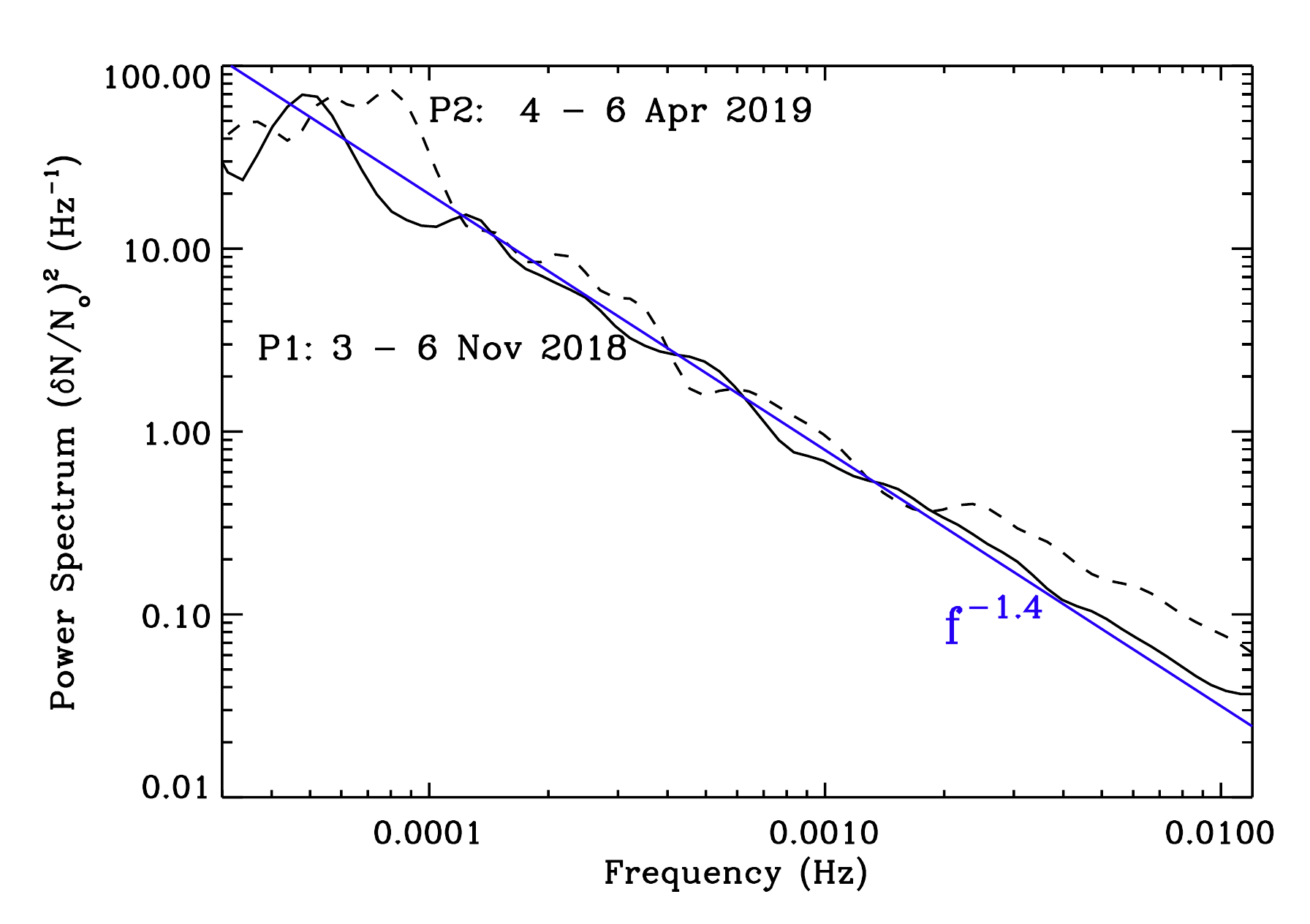}
       \caption{\label{turbulence} Normalized power spectrum of density fluctuations obtained at $\sim$ 36 $R_\sun$, for both encounters E01 (solid line) and E02 (dashed line).  The blue solid line shows the corresponding power law spectrum, varying as $f^{-1.4}$}
\end{figure}

The 7-sec resolution of the density measurements obtained by the QTN analysis on PSP enables us to deduce the spectrum of compressive fluctuations at $\sim$36 $R_\sun$, during the two first encounters. To analyse the electron density fluctuations, we use the Morlet wavelet transforms, which are convenient to unfold turbulence signals into both space (or time) and scale \citep{farge92}. Figure \ref{turbulence} shows the spectrum of the electron density fluctuations, normalized to the corresponding mean density value, for each perihelion where the radial distance is almost constant, close to 0.17 AU. Spectra vary as a $f^{-1.4}$  power-law (blue solid line) in the  $10^{-4}$ - $10^{-2}$ Hz frequency range, in agreement with some previous analyses in this frequency range at 1 AU \citep{int75, iss10, rob17}. For  E01, from 3 to 6 November 2018, the density measurements (see Figures \ref{nTc} and \ref{histogramsN}) of mean  value $\simeq$ 370 cm$^{-3}$  suggest a low speed wind stream \citep{kasp19}. In contrast, for E02, from 4 to 6 April 2019, the plasma is more dilute with a mean density  $\simeq$ 160 cm$^{-3}$ (see section \ref{sec:histograms}). Although, PSP explores different types of wind in encounters 1 and  2, the corresponding amplitude of the normalized power spectrum of the  electron density is similar, around 1 at $10^{-3}$ Hz. However, at higher frequencies, above 2 $10^{-3}$ Hz, the amplitude of the spectrum is { twice higher for E02 than for E01. We have checked that the enhancement of this part of the spectrum corresponds to the electron density structure of day 5 April 2019 (see Figure \ref{nTc}), coming from the fastest solar wind stream, with  alfvenic turbulent fluctuations \citep{kasp19}}. Hence, this dilute solar wind has a higher level of turbulence fluctuations.

A detailed study of the 3D structure of the turbulence needs in particular to correlate the density spectrum with macroscopic plasma parameters, radial distances, values of the plasma $\beta$, and local magnetic field switch back fluctuations, which are outside the scope of this paper.
As it approaches closer to the Sun, Parker Solar Probe will give new crucial clues.
\section{Concluding remarks} 

For these preliminary results on far solar encounters, QTN spectroscopy has been implemented in a simplified way, mainly due to the small antenna length, whereas the suprathermal electrons have been considered globally, neglecting their anisotropy. These simplifications did not affect the accuracy of the density measurements;  concerning the temperatures, the present preliminary determinations of the radial profiles are expected to be more robust than the absolute values.

For both studied encounters, the radial profile of the temperature $T_c$ of the thermal component varies as $R^{-0.74}$  with similar values at equal distances. These values are in the range of Helios measurements. The temperature $T_h$ of the suprathermal component has a very weak radial variation, as expected from its virtually collisionless state, with extrapolated values at the corona compatible with previous coronal measurements close to the solar equator.
As a by-product, the accurate determination of the electron density by the QTN enables us to deduce the low-frequency spectrum of compressive fluctuations around perihelion, varying as f$^{-1.4}$.

Future works will require a detailed study of the structure of the wind, using the parameters available, especially the magnetic field and the velocity. In particular, we intend to study the electron density and temperature(s) behavior during the magnetic field reversals and jumps in speed \citep{bal19}, and to explore the relation between density and core temperature with radial distance. 
We will  also exploit the measured electron temperature gradient to study the interplanetary potential and the heat transport. 

It is noteworthy that the present results have been obtained independently of those from the inboard electron analyzer of the SWEAP instrument suite \citep{kas16}. Comparisons between the latter results and those from QTN spectroscopy should benefit to both techniques. Concerning QTN spectroscopy, we intend to model the suprathermal electrons by taking into account the anisotropy of the Strahl component, and to implement the technique via fitting the whole QTN spectrum, when PSP will be close enough to the Sun for the Debye length to be smaller than the antenna length.
\acknowledgments
\subparagraph{ Acknowledgments} 
Parker Solar Probe was designed, built, and is now operated by the Johns Hopkins Applied Physics Laboratory as part of NASA's Living with a Star (LWS) program (contract NNN06AA01C). Support from the LWS management and technical team has played a critical role in the success of the Parker Solar Probe mission.
We warmly congratulate all the scientists, technicians, engineers and administrators who contributed to this outstanding mission. 

This work is based on observations with the FIELDS instrument suite embarked on Parker Solar Probe whose data are publicly available at http://fields.ssl.berkeley.edu/data/.
S.D. Bale acknowledges the support of the Leverhulme Trust Visiting Professorship program.
In France, this work was supported by CNES and by CNRS/INSU. We thank O. Alexandrova for helpful discussions. 
\appendix
{
\section{Antenna arms separation}
The response $F(kL)$ of a dipole made of two collinear wires of length L with a gap of length $s$ between the wires reads:  (from Eq.(32) by \citet{mey89})
\begin{eqnarray}
F(kL) =  \frac{J_0^2(ka)}{2k^2L^2} \{ && k(L+s) \;{\rm Si}(k(L+s))+kL \;{\rm Si}(kL)  \nonumber  \\ 
&-& k(L+s/2) \;{\rm Si}(k(2L+s))-ks/2\;{\rm Si}(ks)     \\
&-& 4  \sin^2 (k(L+s)/2) \sin^2(kL/2) \;\;\;\;\; \}   \nonumber 
\label{FkL}
\end{eqnarray}
(Si denotes the sine integral function and $J_0$ the Bessel function of order 0). 
In particular, from the asymptotic values of $F(kL)$, an equivalent dipole wire length may be defined, being $(L+s)$ for $ kL \rightarrow 0 $ and $L$ for $kL  \rightarrow \infty$. 

The QTN plateau level $V^2_0$ given in Eq.(\ref{v2min}) depends on the response $F(kL)$, including the separation $s$, via the integral over wavenumbers $k$, noted $F_0$ given by : 
\begin{equation}
F_0 =   \int_0^\infty \frac{F(kL) k L_D^2 }{[k^2L_D^2+1]^2}dk 
\label{F0}
\end{equation}
Hence the effect on the QTN plateau level may be calculated as a function of the dimensionless ratios $L/L_D$ and $s/L$  at each step of the algorithm used to derive the thermal electron temperature. 
When using the asymptotic values of $F(kL)$ sketched above, it is noteworthy that the effect of the wire separation on the QTN plateau level is equivalent to that produced by a wire length of $(L+s)$ for $L_D \gg L$, whereas for $L_D \ll L $ (which may happen closer to the Sun), the separation effect will be negligible.
The function $F_0$ is shown in Fig.~\ref{F0gap} as a function of $L/L_D$ for different values of the separation in the range $[0-3]$m. 
 
On PSP, the physical separation between the antenna wires V1 and V2 is  $\sim 3$ m, but is alleviated by the presence of the spacecraft, so that the effective separation is much smaller. By fitting  $s$  in the range $[0-3]$m with the algorithm used to solve the implicit Eq.(\ref{v2min}), we have empirically determined $s \sim 1.5$ m, so that, for large $L_D$ and for frequencies below $f_p$, V1-V2 behaves as  a dipole of $2 \times 3.5$ m. 

\begin{figure}[ht!]
	\centering
	\includegraphics[scale=0.5]{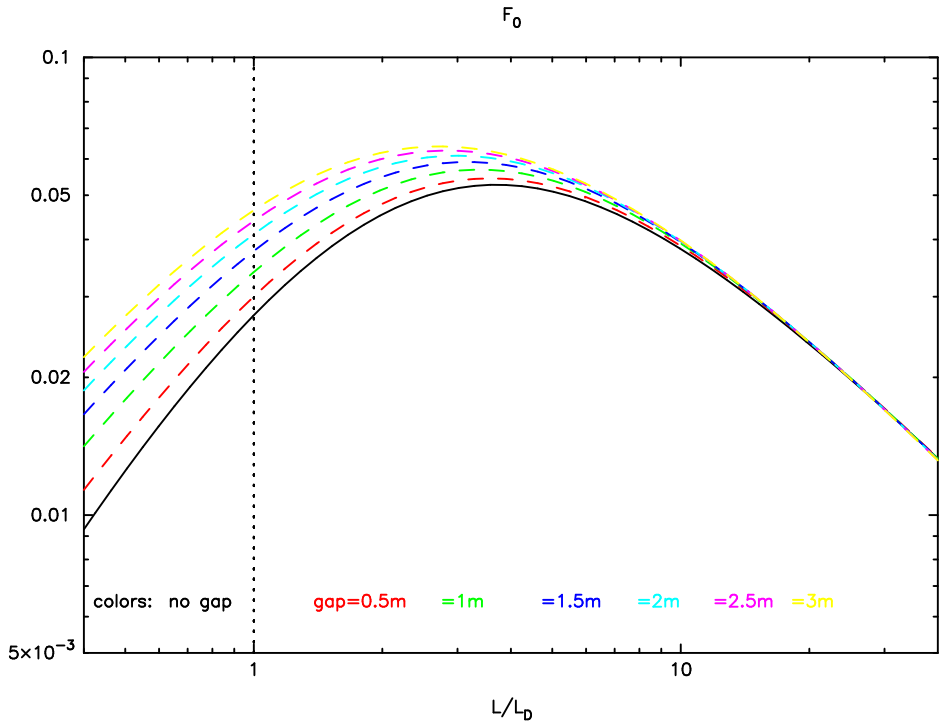}
	\caption{\label{F0gap} Effect of the separation of the antenna arms:  QTN plateau in V$^2/$Hz with $T_c$ in K, normalized to $ 8.14 \times 10^{-16} \sqrt{T_c}$  as a function of $L/L_D$ with separation (dashed color lines) and without separation (continuous black line).}
\end{figure}
}
{
\section{The simplified QTN method and other noises} 
Figure \ref{gudule} shows an example of implementation of the simplified QTN method (without fitting) to obtain the preliminary results reported in the present paper, based on the plasma peak $f_p$ and the minimum $V^2_{min} $ and maximum $V^2_{max} $ noise levels. 
\begin{figure}[ht!]
\centering
 \includegraphics[scale=0.46]{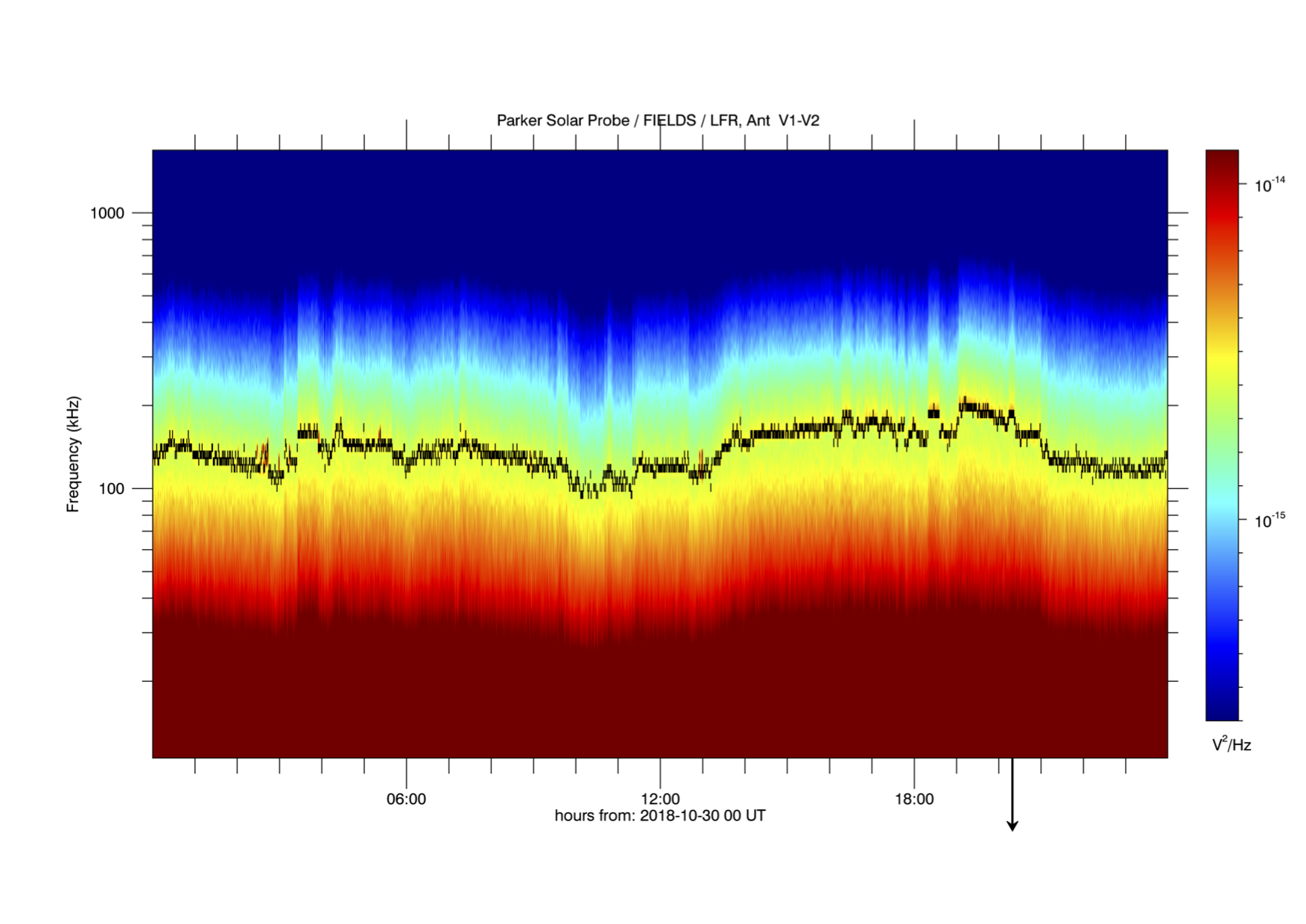}
 
 \vspace*{-63mm}
 \includegraphics[scale=0.7]{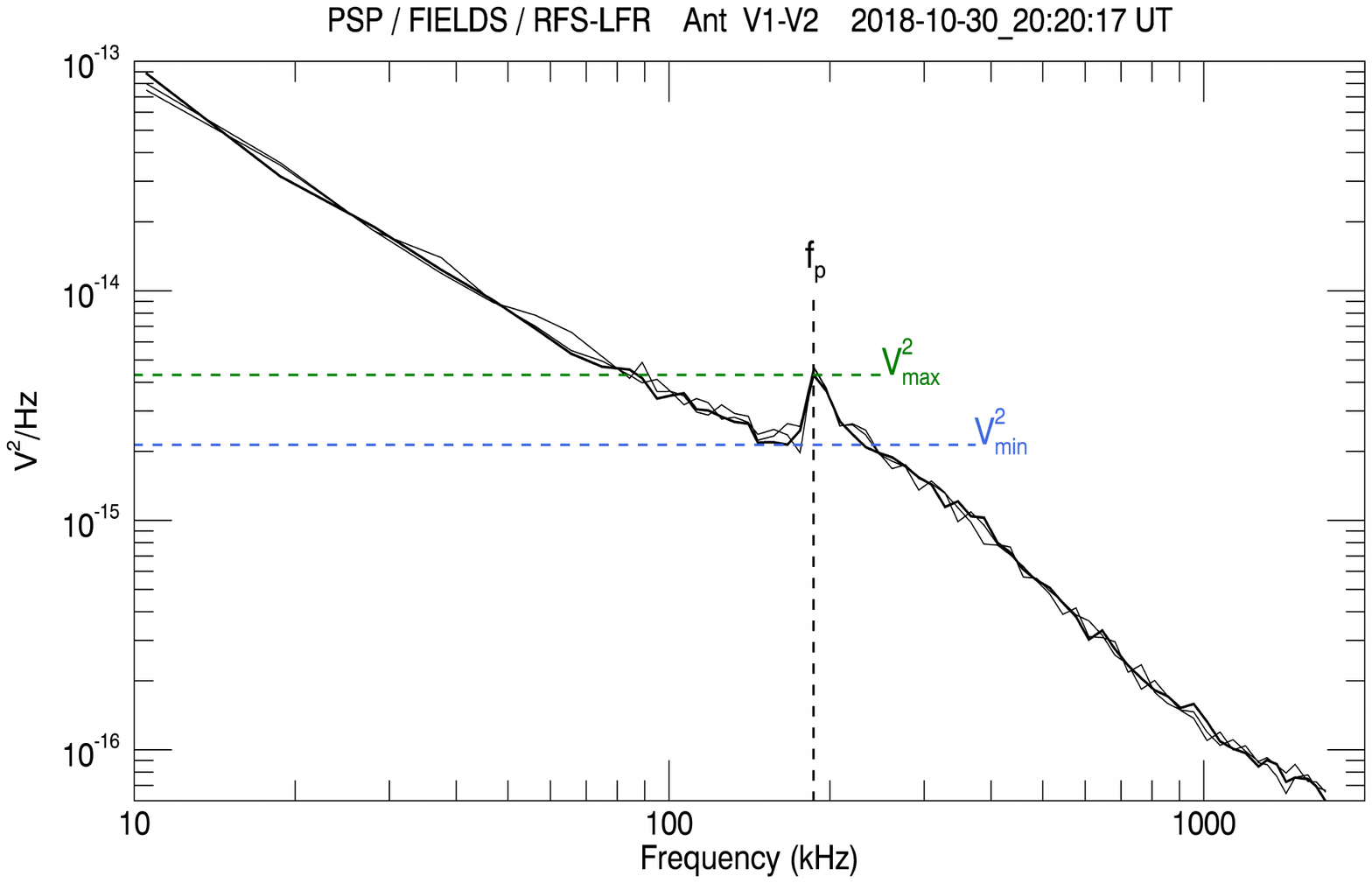}
 
 \vspace*{-60mm}
\caption{\label{gudule} Top panel: Example of a daily spectrogram (2018/10/30, 57s acquisition cadence) in the LFR-RFS frequency range, with the detected plasma peaks $f_p$ superimposed as black bars, which enables us to deduce the electron density. Bottom panel: example of 3 consecutive calibrated spectra (middle acquisition UT indicated on the top, and by the arrow from the spectrogram time line), showing $f_p$ and the two noise levels (dashed lines) used in this paper to determine the thermal and suprathermal temperatures.}
\end{figure}
\clearpage

\paragraph{Dealing with shot and proton noises} 
 As indicated in section 3.2, the observed QTN minimum below $f_p$ reads  $V^2_{min} = V^2_{0} + V^2_{shot} + V^2_{p}$, where
the main contribution is the QTN plateau $V^2_{0}$ given by Eq.(\ref{v2min}).  
We focus here on the minor contributions arising from the shot noise $V^2_{shot}$ and the Doppler-shifted proton noise $V^2_{p}$, which are both varying as $\sqrt{T_c}$ but whose variation depends on unknown parameters which are the antenna d.c. potential $\phi$ and mainly the wind bulk velocity, respectively :  

1) the shot noise, including photoelectron, plasma and bias currents, is detailed in \citet{mey17}, with estimations for PSP in the case $L_D \ll L$. Our simplified method computes it for any $L_D$ (from Eq.(2) by \cite{mon05}), but assuming $\phi=0$  and that antenna biasing yields a stable noise modifying $V^2_{shot}$ by a fixed factor. The ratio $V^2_{shot}/ V^2_{0}$ has been checked to be less  than 10\% for both encounters. Note that  we have kept all the available spectra  for the first encounter, whereas for the second encounter,  we have withdrawn the short (half-hour) periodic tests of unbiasing. That explains why, for the first encounter, we can see some periodic short (and unphysical) drops of  $T_c$ values in Figure \ref{nTc} (top panel).

2) the QTN $V^2_p$ produced by the plasma protons, Doppler-shifted by the solar wind speed, and especially its variation with $\sqrt{T_c}$ for different values of $L/L_D$ is given by \citet{iss99b} (Eq.(22) and Fig.~1 therein). With a wind speed varying from 200 to 400 km/s, and $L/L_D$ between 0.5 and 2 (i.e. typical ranges for both encounters), this yields a ratio $r=V^2_p/V^2_0$ less than 5\% within these ranges. So, for each spectrum, we fit $r$ in the range $[0-0.05]$ in our algorithm used to determine $T_c$.  Finally note that $V^2_p$ is expected to be negligible as PSP will approach the Sun (\citet{mey17}, section 2.8).

}




\end{document}